%% file: document.tex
\documentclass[sigconf, screen]{acmart}

\AtBeginDocument{%
  \providecommand\BibTeX{{%
    Bib\TeX}}}

\usepackage{graphicx}
\usepackage{textcomp}
\usepackage{xcolor}
\usepackage{graphicx}
\usepackage{subfiles}
\usepackage{enumitem}
\usepackage{adjustbox}
\usepackage{float}
\usepackage{tabularx}
\usepackage{multirow}
\usepackage{caption}
\usepackage{amsmath,amsfonts}
\usepackage{url}
\usepackage{booktabs}
\usepackage{siunitx}
\usepackage{fancyvrb}
\usepackage{rotating}
\usepackage{array}
\usepackage{collcell}
\usepackage[many]{tcolorbox}
\usepackage{makecell}
\usepackage{soul}
\usepackage{listings}
\usepackage{balance}
\usepackage{algorithm}
\usepackage{algpseudocode}
\theoremstyle{definition}
\newtheorem{defn}{Definition}
\usepackage[utf8]{inputenc}
\usepackage{microtype}

\newtcolorbox{boxK} {
    sharpish corners,
    boxrule = 0pt,
    toprule = 3pt,
    enhanced,
    fuzzy shadow = {0pt}{-2pt}{-0.5pt}{0.5pt}{black!35},
    boxsep=0mm,
    left=2mm,
    right=2mm,
    top=1.5mm,
    bottom=1.5mm
}

\definecolor{diffgreen}{rgb}{0.1,0.7,0.2}
\lstdefinestyle{javastyle}{
    language=java,
    basicstyle=\ttfamily\footnotesize,
    breaklines=true,
    commentstyle=\color{gray},
    keywordstyle=\color{blue},
    stringstyle=\color{orange},
    showspaces=false,
    showstringspaces=false,
    showtabs=false,
    tabsize=2,
    frame=none,
    xleftmargin=1mm,
    moredelim=[is][\color{diffgreen}]{@g@}{@g@}
}

\definecolor{commandblue}{rgb}{0.2,0.4,0.8}
\definecolor{paramgreen}{rgb}{0.2,0.6,0.2}
\lstdefinestyle{bashstyle}{
    language=bash,
    basicstyle=\ttfamily\footnotesize,
    breaklines=true,
    commentstyle=\color{gray},
    keywordstyle=\color{commandblue},
    stringstyle=\color{paramgreen},
    showspaces=false,
    showstringspaces=false,
    showtabs=false,
    tabsize=2,
    frame=none,
    xleftmargin=1mm,
    moredelim=[is][\color{commandblue}]{@c@}{@c@},
    moredelim=[is][\color{paramgreen}]{@p@}{@p@}
}

\newcommand{\customcell}[1]{%
  \begin{tabular}[t]{@{}l@{}r@{}}
    #1
  \end{tabular}%
}

\newcolumntype{C}{>{\collectcell\customcell}c<{\endcollectcell}}

\def\BibTeX{{\rm B\kern-.05em{\sc i\kern-.025em b}\kern-.08em
    T\kern-.1667em\lower.7ex\hbox{E}\kern-.125emX}}

\setcopyright{cc}
\setcctype{by}
\acmDOI{10.1145/3832783.3834426}
\acmYear{2026}
\copyrightyear{2026}
\acmISBN{979-8-4007-2882-2/2026/10}
\acmConference[ASE '26]{Proceedings of the 41st IEEE/ACM International Conference on Automated Software Engineering}{October 12--16, 2026}{Munich, Germany}
\acmBooktitle{Proceedings of the 41st IEEE/ACM International Conference on Automated Software Engineering (ASE '26), October 12--16, 2026, Munich, Germany}
\acmSubmissionID{ase26main-p2896-p}
\received{2026-03-26}
\received[accepted]{2026-06-18}

\begin{document}

\title{Discovering Performance Archetypes: Critical-Path-Aware Pattern Analysis and Regression Detection}

\author{Kaveh Shahedi}
\orcid{0009-0001-4018-5113}
\affiliation{%
  \institution{Polytechnique Montréal}
  \department{Computer Engineering and Software Engineering}
  \city{Montréal}
  \country{Canada}
}
\email{kaveh.shahedi@polymtl.ca}

\author{Heng Li}
\correspondingauthor
\orcid{0000-0001-5441-6763}
\affiliation{%
  \institution{Polytechnique Montréal}
  \department{Computer Engineering and Software Engineering}
  \city{Montréal}
  \country{Canada}
}
\email{heng.li@polymtl.ca}

\author{Maxime Lamothe}
\orcid{0000-0003-3705-6238}
\affiliation{%
  \institution{Polytechnique Montréal}
  \department{Computer Engineering and Software Engineering}
  \city{Montréal}
  \country{Canada}
}
\email{maxime.lamothe@polymtl.ca}

\author{Foutse Khomh}
\orcid{0000-0002-5704-4173}
\affiliation{%
  \institution{Polytechnique Montréal}
  \department{Computer Engineering and Software Engineering}
  \city{Montréal}
  \country{Canada}
}
\email{foutse.khomh@polymtl.ca}

\subfile{sections/abstract}




\maketitle

\subfile{sections/introduction}

\subfile{sections/related_works}

\subfile{sections/methodology}

\subfile{sections/results}

\subfile{sections/threats}

\subfile{sections/conclusion}

\subfile{sections/data_availability}

\subfile{sections/acknowledgement}

\clearpage

\balance
\bibliographystyle{ACM-Reference-Format}
\bibliography{document}

\end{document}

%% file: sections/abstract.tex
\begin{abstract}

Software performance analysis and prediction requires integrating multiple signals, as code structure alone cannot capture runtime behavior shaped by execution frequency, resource contention, and I/O patterns. We present a \textit{critical-path-aware performance analysis} methodology that automatically discovers recurring performance patterns by synthesizing static code features, dynamic execution traces, and kernel-level resource data. In a preliminary study across six real-world C/C++ applications (SQLite, OpenSSL, Zstandard, FFmpeg, cURL, and jq), we first empirically confirm that static complexity metrics explain only 10.4\% of the variance ($\rho^2$) in critical path execution time, quantifying a gap that, while theoretically expected, had not been measured systematically across applications. Motivated by this finding, we analyze nearly 80,000 critical execution paths and address two research questions. First, we discover 13 distinct \emph{performance archetypes}: recurring behavioral patterns that appear consistently across different applications, independent of their domain or implementation. Five of these patterns are near-universal and appear in at least five of the six applications studied. Notably, three of these archetypes are present in all six applications, and together, these common patterns account for 56.4\% of all observed paths. Each archetype maps to specific resource profiles and optimization strategies that transfer across domains. Second, we leverage these archetypes within a multi-signal regression detection framework that triangulates path structure, resource consumption, and archetype deviations, achieving an F1-score of 0.867 and a 60.4\% improvement over resource-only methods.

\end{abstract}

%% file: sections/introduction.tex
\section{Introduction}
Software performance remains a critical concern in modern computing systems, where even minor inefficiencies can cascade into significant resource waste and degraded user experiences~\cite{jin2012understanding, zaman2012qualitative, zhao2022large, liu2014characterizing}. As applications grow in complexity and scale, understanding and predicting their performance characteristics becomes increasingly challenging. Traditional performance analysis approaches typically rely on either static code metrics~\cite{balsamo2004model, menzies2007datamining, shin2011evaluating} or dynamic runtime profiling~\cite{balsamo2004model, ball1996efficient}, each offering complementary but incomplete perspectives on system behavior.

A fundamental challenge is that static analysis cannot fully predict runtime performance. This limitation is theoretically well-understood: static analysis cannot determine loop iteration counts (a potentially undecidable problem), does not capture the cost of individual instructions or library calls, and cannot know execution frequency for a given input~\cite{trubiani2011detection, menzies2010defect, nunez2017source, velez2020relation}. Conversely, dynamic profiling captures real execution behavior but typically focuses on single signals such as CPU time or function call counts, missing the complex relationships between computation, memory access patterns, and I/O operations that characterize modern application performance~\cite{shahedi2024tracing, vanhoorn2012kieker, liao2020using, eismann2020microservices}. While practitioners may intuit this disconnect, no prior work has quantified it systematically across multiple applications or provided a constructive response.

This gap motivates three challenges that our methodology addresses. First, \textit{discovering recurring performance patterns that transfer across applications}, as existing performance analysis tools lack models capturing recurring execution patterns across diverse workloads~\cite{muhlbauer2023analyzing, guo2016empirical, pereira2020sampling, lesoil2023learning, jamshidi2017transfer}. Second, \textit{detecting regressions by triangulating multiple signals}, since single-signal approaches suffer from high false positive rates~\cite{eismann2020microservices, nguyen2012automated, shang2015automated, huang2014performance, chen2020perfjit}. Third, \textit{quantifying which functions truly matter for performance} rather than relying on static proxies, since developers frequently invest effort optimizing statically complex functions that rarely impact runtime performance~\cite{zhao2022large, selakovic2016performance, nistor2015caramel, linares2015developers}.

We present a comprehensive methodology that synthesizes static code analysis, dynamic critical path extraction, and kernel-level resource monitoring into an automated pipeline. Our approach analyzes the top-$k$ critical execution paths (i.e., sequences of function calls that contribute most significantly to overall wall-clock execution time) extracted from multiple executions across diverse inputs. By studying these critical paths across six real-world applications (\textit{SQLite}, \textit{OpenSSL}, \textit{Zstandard}, \textit{FFmpeg}, \textit{cURL}, and \textit{jq}), we concentrate analysis on code segments that truly impact performance and enable cross-application pattern discovery. This work addresses two research questions:

\textbf{RQ1:} \textit{Can critical paths be automatically clustered into transferable performance archetypes?} We explore whether critical paths exhibit recurring multi-dimensional patterns that transcend application boundaries and enable cross-application performance characterization and optimization guidance.

\textbf{RQ2:} \textit{Can multi-signal triangulation improve automated performance regression detection?} We examine whether combining path structure, resource consumption, and archetype deviations enables more reliable and automated regression detection compared to single-signal approaches.

Our contributions advance the state-of-the-art in automated performance analysis:
(1) \textbf{An empirical quantification of the static-dynamic gap} across six real-world applications, showing that static complexity metrics explain only 10.4\% of performance variance and 14.7\% of functions exhibit misaligned behavior, motivating the need for dynamic approaches (preliminary study);
(2) \textbf{A novel archetype discovery methodology} that automatically extracts multi-dimensional feature vectors from execution traces and discovers 13 recurring performance archetypes through clustering, with five near-universal patterns appearing across at least five of six applications (three of which appear in all six), collectively covering 56.4\% of observed paths, enabling cross-application performance characterization;
(3) \textbf{A triangulated regression detection framework} that synthesizes path structure, resource consumption, and archetype deviations to detect performance anomalies with F1-score of 0.867 and 60.4\% improvement over resource-only methods, while providing actionable root cause attribution;
and (4) \textbf{An open-source, end-to-end automated pipeline} from trace collection to archetype discovery and regression detection, enabling reproducibility and facilitating adoption in automated performance engineering workflows.

%% file: sections/related_works.tex
\section{Related Works}
\label{sec:related-works}

\subsection{Software Performance Analysis}

Understanding and predicting software performance is a crucial challenge in software engineering, with a persistent disconnect between static code structures and runtime behaviors.

\textbf{Performance Regression Detection.}
Detecting performance regressions remains a challenge. Early statistical approaches applied statistical process control~\cite{nguyen2012automated, malik2013automatic, jiang2009automated} but suffered from false positives, leading to machine learning solutions~\cite{chen2020perfjit, chen2017exploratory, he2019statistics, shahedi2024tracing, liu2008isolation, zhang2019robust, scholkopf2001estimating}. However, industrial deployments show limitations~\cite{foo2015industrial, liao2025early, reichelt2019peass}. For example, approaches may detect regressions but fail to pinpoint code causes~\cite{foo2015industrial}, identify only architectural (not function-level) bottlenecks~\cite{liao2025early}, or capture only local unit test effects~\cite{reichelt2019peass}. Notably, Chen and Shang~\cite{chen2017exploratory} observed that seemingly innocuous changes, such as small modifications to configurations or data structures, often cause major regressions, highlighting static analysis limitations. Recent work has broadened the scope: HybridRCA~\cite{ekhlasi2025hybridrca} uses critical-path extraction and targeted metric collection for root-cause analysis in microservices; JPerfEvo~\cite{shahedi2025jperfevo} detects method-level performance changes per commit in Java projects; and Zhao et al.~\cite{zhao2024performance} predict performance bugs from source code metrics. All three require code-change context or source-level analysis, differing from our execution-level, diff-free regression detection setting. \textit{Existing approaches mostly rely on single signals in isolation and version-level comparisons, leaving open how regressions can be detected at individual execution level using multi-dimensional behavioral signals.}

\textbf{Performance Characterization and Patterns.}
Foundational work by Smith~\cite{smith1990performance, smith2001performance} introduced execution graphs and software performance engineering principles, establishing that performance must be designed in from the start. Building on this, researchers have sought to characterize performance via workload modeling, evolving from early user behavior patterns~\cite{menascé1999customer, menascé2001capacity} to sophisticated session-based mining~\cite{vögele2018wessbas, vögele2015automatic}. As architectures shifted, new challenges emerged; for instance, Eismann et al.~\cite{eismann2020microservices} demonstrated that single execution runs yield unreliable measurements due to high variance. Consequently, even careful experimental designs~\cite{heinrich2017performance, calzarossa2016workload}, clustering~\cite{avritzer2002software}, or forecasting~\cite{schulz2019context} methods focus on external workload characterization (what users do) rather than internal execution behaviors (how systems respond). Industry APM tools (Datadog, New Relic, Dynatrace) monitor performance at the service or endpoint level, but do not provide function-call-granularity analysis or automated pattern discovery across workloads. \textit{Existing approaches describe system inputs rather than recurring internal performance behaviors, leaving open whether execution paths exhibit predictable, transferable patterns.}

\textbf{The Static-Dynamic Divide.}
A visible tension exists between static analysis-based predictions and runtime reality. While static complexity metrics~\cite{mccabe1976complexity} were studied for defect prediction~\cite{basili1996validation, shepperd1988critique}, dynamic analysis emerged to capture actual runtime behavior, revealing program invariants and anomalies invisible to static analysis~\cite{ernst2001dynamically, mock2003dynamic}. These two worlds remain largely separate. Even recent machine learning approaches for performance prediction~\cite{syer2013leveraging, cortellessa2014approach}, anomaly detection~\cite{nandi2016anomaly, soldani2022anomaly}, or pattern mining~\cite{lo2008mining, han2012mining} typically operate on either static or dynamic features, not both. \textit{No prior work has systematically quantified how severely static complexity fails to predict runtime criticality, nor leveraged dynamic patterns for transferable performance characterization.}

\subsection{Software Instrumentation and Tracing}

Performance analysis fundamentally depends on observing system behavior, yet instrumentation choices profoundly shape what can be discovered.

\textbf{The Multi-Level Tracing Landscape.}
Modern analysis requires synchronizing application and system-level visibility. Tracing tools like LTTng~\cite{desnoyers2006lttng} offer low-overhead (<$1\%$) kernel and user-space tracing, making it suitable for production use at companies like Google, IBM, and Ericsson. This is complemented by tools like uftrace~\cite{kim2017uftrace}, perf~\cite{de2010new}, and eBPF~\cite{gregg2019bpf}. These tools present fundamental trade-offs between overhead, detail, and safety~\cite{gregg2015choosing, prasad2005locating, cantrill2004dynamic}. \textit{Prior work has rarely leveraged synchronized multi-level traces correlating code execution with resource consumption.}

\textbf{Critical Path Analysis: From Parallelism to Sequential Code.}
Critical path analysis, which identifies the longest dependency chain, originated in parallel computing~\cite{yang1988critical} and was extended to distributed systems~\cite{denys2023distributed}. This concept remains valuable in modern microservices, as exemplified by Kaldor et al.~\cite{kaldor2017canopy} who analyzed RPC-level paths across ~40,000 endpoints at Uber, and in cyber-physical systems~\cite{hendriks2017analyzing}. These methods find bottlenecks missed by aggregate metrics. \textit{All prior approaches focus on systems with explicit parallelism; no work has extracted critical paths at function-call granularity within dynamic call graphs for automated pattern discovery across applications.}

\textbf{Trace Analysis: Visualization Versus Understanding.}
Extracting insights from traces requires sophisticated analysis. Tools like Eclipse Trace Compass~\cite{trace2025compass} provide comprehensive analysis with Control Flow views, Resource views, and CPU Usage analysis with flame graphs, while Jaeger~\cite{jaeger2023tracing} is standard for distributed tracing. While some research addresses event correlation~\cite{poirier2010accurate, zhao2017correlating, binder2009portable}, \textit{Existing analysis remains visualization-focused; no prior work applies machine learning to automatically extract recurring performance archetypes from multi-dimensional trace data.}

%% file: sections/methodology.tex
\section{Methodology}
\label{sec:methodology}

\begin{figure*}[!t]
\centering
\includegraphics[width=0.9\textwidth]{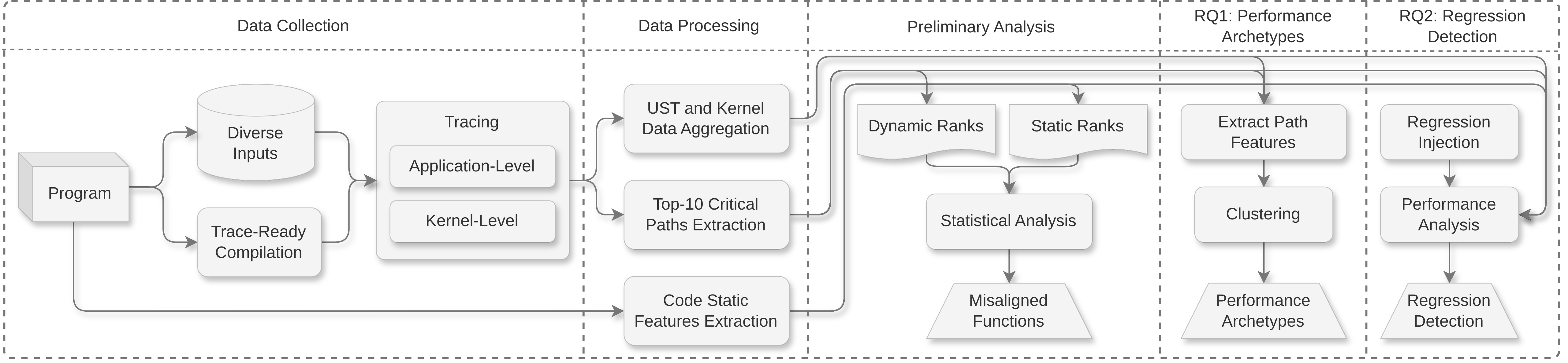}
\captionsetup{justification=centering}
\caption{The overview of our study.}
\label{fig:overview}
\vspace{-3mm}
\end{figure*}

\subsection{Overview}

Traditional performance analysis approaches often examine code structure (static analysis) or execution time (dynamic profiling) in isolation, potentially overlooking crucial interactions with system resources. To address this limitation, we propose a comprehensive, multi-layered framework that integrates three complementary perspectives for each program execution:
\begin{itemize}[topsep=2pt,itemsep=0pt,parsep=0pt,leftmargin=*]
  \item \textbf{Static Code Characteristics:} Structural properties extracted directly from source code that indicate potential complexity.
  \item \textbf{Dynamic Execution Traces:} Fine-grained application-level function call graphs captured during runtime.
  \item \textbf{Kernel-Level Resource Monitoring:} System-wide CPU, memory, and disk I/O events observed at the operating system level.
\end{itemize}

By correlating these three perspectives, our methodology enables a deeper understanding of performance phenomena that would be invisible to any single-perspective analysis approach. Specifically, we identify recurring, resource-coupled performance patterns (i.e. ``archetypes'') across diverse applications (RQ1) and develop a robust technique for detecting performance regressions through performance modeling using multiple signals (RQ2). Figure~\ref{fig:overview} illustrates the overview of this work.

\textbf{Terminology.} Throughout this paper, we measure performance primarily as \emph{wall-clock execution time (i.e., latency)}, complemented by CPU utilization, memory allocation rate, and I/O throughput as secondary metrics. When we refer to ``performance variance,'' we mean the explained variance ($\rho^2$) of execution time.

\subsection{Background: Critical Paths and Archetypes}
\label{sec:background}

Before detailing our methodology, we explain the two core concepts that underpin our approach and motivate the research questions.

\textbf{What is a critical path?} In the context of this work, a \emph{critical path} is the sequence of function calls from root to leaf in a dynamic call graph that contributes the most wall-clock time to an execution. Unlike aggregate profiling (which ranks functions by cumulative time across many invocations), critical paths preserve the \emph{calling context}: they show not just which functions are expensive, but how those functions are reached through specific caller chains. A single application execution may contain thousands of function calls, but only a small number of root-to-leaf paths dominate the total execution time. By extracting the top-$k$ such paths per execution, we reduce a complex call graph to a compact set of performance-relevant sequences.

\textbf{Why do critical paths matter?} Critical paths are valuable because they (1) directly identify the code sequences responsible for the majority of execution time, where optimizing on-path functions yields the highest return; (2) carry multi-dimensional information, such as structural (e.g., depth, function sequence), temporal (e.g., duration, time distribution), and resource (e.g., CPU, memory, I/O), which are richer than single-metric profiles; and (3) enable cross-input pattern analysis by capturing how performance behavior varies with workload.

\textbf{Why cluster critical paths into archetypes?} Clustering critical paths based on their multi-dimensional features reveals recurring patterns across applications, which we call \emph{performance archetypes}. Each archetype represents a distinct mode of execution (e.g., ``memory-intensive compression,'' ``fast initialization'') with specific resource profiles and optimization strategies. Archetypes also serve as a basis for anomaly detection: if an execution suddenly shifts from one archetype to another, this may indicate a performance deviation.

\subsection{Subject Programs and Data Collection}

To evaluate our methodology, we select six widely-used, open-source C/C++ applications representing diverse computational domains and exhibiting different performance characteristics (Table~\ref{tab:subject_systems}). We obtained stable release versions of each application from their official repositories and websites. These applications were chosen to provide broad coverage of common performance behaviors:
\begin{itemize}[topsep=2pt,itemsep=0pt,parsep=0pt,leftmargin=*]
  \item \textbf{SQLite:} A database library that alternates between I/O-bound (disk reads/writes) and compute-bound (query processing) phases.
  \item \textbf{OpenSSL:} A cryptography toolkit that is primarily CPU-intensive due to complex mathematical operations.
  \item \textbf{Zstandard (zstd):} A compression utility that balances significant CPU usage, memory management for buffering, and disk utilization for storage.
  \item \textbf{FFmpeg:} A multimedia framework characterized by streaming I/O combined with parallel processing for encoding/decoding.
  \item \textbf{cURL:} A network transfer tool that is primarily I/O-bound due to network latency, DNS resolution, and TLS handshake operations, with intermittent CPU bursts for protocol parsing.
  \item \textbf{jq:} A lightweight command-line text processor that is primarily CPU-bound, performing recursive tree traversal, pattern matching, and data transformation on structured input.
\end{itemize}

As shown in Table~\ref{tab:subject_systems}, these programs vary considerably in size and complexity, providing a rich dataset for evaluating the generalizability of our findings. For each application, we generated 500 distinct inputs designed to trigger diverse execution paths and workload characteristics.

\textbf{Workload generation strategy.}
We use \emph{structured randomization}: for each subject program, we enumerate the input parameters from its official documentation (e.g., FFmpeg codec and resolution; SQLite DML/DDL operations and transaction sizes; OpenSSL cipher suites and key-exchange algorithms; zstd compression levels; cURL protocol and option flags; jq filter programs), and then uniformly randomize over their valid ranges to produce 500 distinct inputs. We deliberately do not constrain inputs to pre-known critical code paths, as doing so would bias the discovered archetypes toward already-known behavior and miss emergent patterns. The complete workload generation scripts for each application are provided in our replication package.

Each input was executed three times (i.e., three iterations) to ensure measurement stability and mitigate transient system noise. We chose three iterations after confirming low variance across pilot runs; small iteration counts are accepted practice in rigorous performance evaluation when variance is bounded~\cite{georges2007statistically}. The 500-input target per application balances workload diversity with practical collection cost.

\begin{table*}[h]
\centering
\caption{Subject Applications and their Characteristics}
\label{tab:subject_systems}
\begin{tabular}{lclrrrr}
\toprule
\textbf{Application} & \textbf{Domain} & \textbf{LOC} & \textbf{Avg. Complexity}$^*$ & \textbf{Total Paths} & \textbf{Avg. Length}$^{**}$ & \textbf{Median Length}$^{**}$ \\
\midrule
SQLite & Database & 212K & 6.89 & 15,000 & 11.70 & 12.0 \\
OpenSSL & Cryptography & 632K & 6.08 & 11,088 & 14.70 & 16.0 \\
Zstandard & Compression & 95K & 4.64 & 8,969 & 9.50 & 8.0 \\
FFmpeg & Multimedia & 1.3M & 7.28 & 14,990 & 11.21 & 11.0 \\
cURL & Network Transfer & 197K & 6.81 & 14,975 & 12.55 & 11.0 \\
jq & Text Processing & 115K & 7.63 & 14,767 & 8.42 & 8.0 \\
\bottomrule
\end{tabular}
\\
\vspace{1mm}
\small{$^*$Average Cyclomatic Complexity of functions\\
$^{**}$Average/Median number of functions per critical path}
\\
\vspace{-4mm}
\end{table*}

\subsection{Multi-Level Trace Collection}

For each execution, we collected three synchronized data streams that together provide a comprehensive view of application behavior:

\subsubsection{Static Code Analysis}
We performed automated static analysis on the source code of each application using srcML to extract structural metrics for every function $f$. The collected metrics include: Lines of Code (LOC), Cyclomatic Complexity ($C_{\text{cyclo}}$), maximum Loop Nesting Depth ($N_{\text{nest}}$), number of unique function calls made ($\lvert\text{calls}\rvert$), and a binary indicator for the presence of I/O operations ($\mathbb{I}(\text{has\_io})$). These metrics are combined into a weighted static complexity score $S_{\text{static}}(f)$ (Equation~\ref{eq:static_complexity} in Section~\ref{sec:preliminary-study}), which we use in our preliminary study to quantify the gap between static complexity and dynamic runtime criticality.

\subsubsection{Application-Level Function Tracing}
We instrumented each application with \textit{uftrace}~\cite{kim2017uftrace}, a lightweight user-space instrumentation tool that supports C and C++ programs natively with low overhead and straightforward integration for our use case, to capture detailed function entry and exit events during execution, including precise timestamps and call relationships. Programs were compiled with instrumentation flags (i.e., \texttt{-g -finstrument-functions}) to enable this profiling capability. This provides a complete view of the application's control flow at function-call granularity, revealing which code paths are actually exercised during execution and how much time is spent in each function.

\subsubsection{Kernel-Level Resource Tracing}
Simultaneously with appli-cation-level tracing, we captured fine-grained kernel events with nanosecond precision via LTTng~\cite{desnoyers2006lttng}. We enabled specific event categories relevant to performance analysis:
\begin{itemize}[topsep=2pt,itemsep=0pt,parsep=0pt,leftmargin=*]
  \item \textit{CPU Utilization:} Context switches, process lifecycle events.
  \item \textit{Memory Management:} Page allocations and deallocations at both kernel and user-space levels.
  \item \textit{Disk I/O:} Block-level read/write requests and completions.
  \item \textit{System Calls \& Network:} System call entry/exit and network device events.
\end{itemize}

Process and thread identifiers were included to correlate kernel events with application-level traces. User-space memory allocation events (e.g., \texttt{malloc}, \texttt{free}) were captured through LTTng's LD\_PRELOAD-based library interposition mechanism to instrument UST (User Space Tracing) memory events. The resulting traces provide a complete picture of how the application interacts with system resources throughout its execution.

\subsection{Critical Paths Extraction and Correlation}
\label{sec:critical-path-extraction}
Central to our analysis is the concept of the application-level \emph{critical path}. We define this concept precisely:

\begin{defn}[Critical Path]
A critical path is the longest call path, in terms of wall-clock time, of a thread during an execution. Formally, given an execution $E$ with function call events organized as a dynamic call tree, the \emph{critical path} $P^* = [f_1, f_2, \ldots, f_n]$ is the root-to-leaf sequence of function invocations such that: (1) each $f_i$ has entry/exit timestamps from the execution trace; (2) $f_1$ is the root call (the entry point of the thread, typically \texttt{main()}); and (3) for each consecutive pair $(f_i, f_{i+1})$, $f_{i+1}$ is the direct callee of $f_i$ with the longest wall-clock duration among all callees of $f_i$.
\end{defn}

% Note that in a call tree, the path duration always equals the root's wall-clock time, since each parent's execution encompasses its descendants.

\textbf{Distinction from classical CPA.} Our definition differs from classical critical path analysis in parallel computing~\cite{yang1988critical}, which identifies the longest \emph{dependency chain across concurrent tasks}. Instead, we greedily descend the call tree by always selecting the longest-duration child at each level, producing the root-to-leaf path through the most time-consuming invocations. The path duration is the root's wall-clock time, since in a call tree each parent's duration encompasses its children. We adopt this definition because our goal is to identify the \emph{function-level sequence} most responsible for execution time, not inter-task dependencies. For multi-threaded executions, we extract critical paths for \emph{each thread} using per-thread timestamps, then report the longest paths across all threads with each path's \texttt{thread\_id} recorded. Thread concurrency therefore affects measured \emph{durations} (due to scheduling and resource sharing) but not path \emph{structure}, since each path is greedily descended within a single thread's call tree.

\textbf{Algorithm.} We extract critical paths using Trace Compass~\cite{trace2025compass} with TMLL~\cite{shahedi2025technical} as the API wrapper. For each of the six applications, we generate 500 distinct inputs and execute each 3 times (iterations), yielding 1,500 executions per application. The extraction algorithm (Algorithm~\ref{alg:critical_path}) operates as follows:

\begin{algorithm}[t]
\small
\caption{Top-$k$ Critical Path Extraction}
\label{alg:critical_path}
\begin{algorithmic}[1]
\Require Call tree $T = (V, E)$ per thread, $k$ (number of paths)
\Ensure Set of $k$ critical paths $\mathcal{P} = \{P_1, \ldots, P_k\}$
\State \textbf{function} \textsc{LongestPath}($v$, excluded\_edges):
\State \quad $\text{best} \leftarrow [v]$; $\text{best\_dur} \leftarrow \text{duration}(v)$
\State \quad \textbf{for each} child $c$ of $v$ where $(v, c) \notin$ excluded\_edges \textbf{do}
\State \quad \quad $(p, d) \leftarrow$ \textsc{LongestPath}($c$, excluded\_edges)
\State \quad \quad \textbf{if} $d > \text{best\_dur}$ \textbf{then}
\State \quad \quad \quad $\text{best} \leftarrow [v] \cup p$; $\text{best\_dur} \leftarrow d$
\State \quad \textbf{return} $(\text{best}, \text{duration}(v))$
\State $P_1 \leftarrow$ \textsc{LongestPath}(root, $\emptyset$) \Comment{Primary critical path}
\For{$i = 2$ to $k$}
    \State $\text{candidates} \leftarrow \emptyset$
    \For{each node $n$ on $P_{i-1}$ with chosen child $c^*$}
        \State $E' \leftarrow \{(n, c^*)\}$ \Comment{Exclude the edge, not the node}
        \State $(p, d) \leftarrow$ \textsc{LongestPath}($n$, $E'$)
        \State Add $(p, d)$ to candidates
    \EndFor
    \State $P_i \leftarrow$ best candidate not in $\{P_1, \ldots, P_{i-1}\}$
\EndFor
\Return $\mathcal{P} = \{P_1, \ldots, P_k\}$
\end{algorithmic}
\end{algorithm}

We set $k=10$ (the number of critical paths to extract per execution, cf.\ Algorithm~\ref{alg:critical_path}) to capture not only the absolute longest path but also significant alternative paths occurring under varying inputs or runtime conditions. We chose $k=10$ because 11.3\% of executions already produce fewer than 10 distinct paths due to limited control-flow complexity; increasing $k$ further yields diminishing returns as the additional paths become near-duplicates of the primary path rather than structurally distinct alternatives. While targeting 10 paths per execution, this 11.3\% subset contributes slightly fewer paths, yielding a total of 79,789 paths across 9,000 executions.

\textbf{Design choice: peak vs.\ aggregate.} Our algorithm selects paths based on the longest single invocation chain, not cumulative time across repeated calls. A function invoked 1,000 times briefly would not appear on the critical path if a single longer invocation exists elsewhere. This is a deliberate design choice: our paths capture \emph{peak} execution behavior rather than \emph{aggregate} impact. The dynamic criticality score $S_{\text{dynamic}}$ (Equation~\ref{eq:dynamic_complexity}) complements this by capturing aggregate importance across all executions. For $k > 1$, the deviation algorithm (lines 9--16 of Algorithm~\ref{alg:critical_path}) finds alternative paths by excluding specific \emph{edges} rather than nodes, preserving intermediate functions that may participate in different call chains.

We correlate application-level function calls with kernel-level resource events by time-aligning the two trace streams. Kernel events are aggregated into $10\,\mu\text{s}$ windows, providing resource usage statistics (i.e., CPU utilization, memory allocation/deallocation activity, and I/O throughput) for each window. Each function on the critical path has a start and end timestamp; we determine which time windows overlap with the function's execution interval and associate the function with the resource consumption patterns observed in those overlapping windows. The $10\,\mu\text{s}$ window is well below typical critical-path function durations (median of hundreds of microseconds), so any sub-window skew is dominated by measurement noise rather than aggregation error. This correlation enables us to understand not just \emph{which} functions are performance-critical, but also \emph{why} (i.e., CPU intensity, memory pressure, or I/O bottlenecks).

\subsection{Discovering Performance Archetypes}

We aim to discover recurring, multi-dimensional performance patterns by clustering critical paths based on their combined structural, temporal, and resource characteristics. Each extracted critical path is represented as a 33-dimensional feature vector comprising five distinct feature categories, which has been similarly done in previous works~\cite{vanhoorn2012kieker, okanovic2016towards}.
\begin{itemize}[topsep=2pt,itemsep=0pt,parsep=0pt,leftmargin=*]
\item \textbf{Structural Features (6D):} Path depth, average cyclomatic complexity of functions along the path, and complexity trend (whether cyclomatic complexity increases, decreases, or remains stable as execution progresses through the path).

\item \textbf{Temporal Features (6D):} Total execution duration, time concentration measured by the Gini coefficient (indicating whether time is evenly distributed across functions or concentrated in a few), and bottleneck characteristics (location and intensity of the most time-consuming segment).

\item \textbf{Resource Features (9D):} Statistical summaries (mean, standard deviation) and behavioral indicators (peak values, growth rates) for CPU utilization, memory allocation activity, and I/O throughput during the path's execution.

\item \textbf{Transition Features (7D):} Entropy and frequency of transitions between function types, where functions are classified based on their dominant resource usage (CPU-bound, Memory-bound, or I/O-bound). High transition entropy indicates frequently changing resource demands.

\item \textbf{Phase Features (5D):} Number and characteristics of resource phase changes within the path (e.g., transitions from CPU-intensive to I/O-intensive). These features capture the dynamic evolution of resource demands along the critical path.
\end{itemize}

After extracting these feature vectors for all critical paths from all distinct executions across all the applications, we normalize them using standard scaling to ensure all features contribute equally to clustering. We then apply $k$-means clustering to group similar paths into performance archetypes. We chose $k$-means for its scalability to our $\sim$80,000-path, 33-dimensional space and its centroid-based output, which directly provides an interpretable prototype for each archetype~\cite{xu2005clustering, song2008trace}. The optimal number of clusters $k^*$ is determined by maximizing the average silhouette score, evaluating candidates in the range $k \in [2, 15]$. We capped the upper bound at 15 to favor parsimony and interpretability: beyond 15 clusters, the additional archetypes begin to split existing patterns rather than revealing new ones, and interpretability is widely recognized as a key criterion for selecting an appropriate cluster count~\cite{hu2026interpretable}.

Each resulting cluster represents a distinct performance archetype, which is a recurring pattern of behavior characterized by specific combinations of structural, temporal, and resource features. We analyze these archetypes to identify their defining characteristics and determine which archetypes are \emph{universal} (appearing in all six applications), \emph{near-universal} (appearing in $\ge$5 of 6 applications), or application-specific. We assess the stability of archetypes by examining their consistency across the execution iterations and measuring intra-cluster homogeneity using the within-cluster sum of squares of feature vector distances to their cluster centroid.

\subsection{Multi-Signal Regression Detection}
\label{sec:regression-injection}

This step involves developing and evaluating a multi-signal approach for detecting performance regressions. To create a realistic evaluation scenario, we simulate regressions by programmatically injecting performance regressions into randomly selected functions within the subject applications. These injections are designed to mimic real-world regression scenarios targeting different system resources, similarly done in previous works~\cite{liao2020using, shang2015automated, ahmed2016studying, shahedi2024tracing}. Table~\ref{tab:regression_types} summarizes the three types of regressions we introduce.

\begin{table}[]
\centering
\caption{Injected Regression Types and Characteristics}
\label{tab:regression_types}
\resizebox{0.9\columnwidth}{!}{
\begin{tabular}{lp{5.7cm}}
\toprule
\textbf{Type} & \textbf{Description} \\ 
\midrule
CPU Bottleneck &
  Adds computational overhead via repeated arithmetic operations, simulating algorithmic inefficiency or unoptimized code paths. \\
\midrule
Memory Bloat &
  Allocates and initializes large memory blocks, simulating memory leaks, excessive buffering, or inefficient data structures. \\
\midrule
I/O Contention &
  Performs repeated file write operations, simulating excessive logging, inefficient I/O patterns, or unnecessary disk access. \\ 
\bottomrule
\end{tabular}%
}
\\
\vspace{-3mm}
\end{table}

For each regression injection, we modify a single function by inserting degradation code at the function's entry point using srcML-based AST manipulation. Target functions are selected from functions that appear on at least one application-level trace in the baseline executions, ensuring the regression has a realistic chance of manifesting. We inject only one regression type per modified build to isolate the effects.

Our detection framework operates in two phases: baseline modeling and anomaly detection. During baseline modeling, we use 70\% of non-regressed executions to establish statistical models of normal behavior. The split is performed \emph{by input} rather than by individual execution: all three iterations of a given input are assigned to the same partition to prevent data leakage. The models capture four dimensions:

\begin{itemize}[topsep=2pt,itemsep=0pt,parsep=0pt,leftmargin=*]
\item \textbf{Path Signature Models:} We characterize typical path structures including path length distributions, common function sequences (using n-grams), and frequently occurring bottleneck functions.

\item \textbf{Resource Distribution Models:} We model the expected statistical properties (mean, standard deviation, percentiles such as p95) of CPU, memory, and I/O usage during normal execution.

\item \textbf{Archetype Distribution Models:} Using the archetypes discovered in RQ1, we compute the probability distribution over archetypes observed during normal execution.

\item \textbf{Performance Bounds Models:} We establish statistical limits (e.g., p99, IQR-based bounds) on critical path durations and other key performance metrics.
\end{itemize}

When evaluating a new execution for potential regressions, we compute four distinct anomaly scores by comparing its features against the corresponding baseline models. Each score is normalized to $[0, 1]$ before aggregation:

\textbf{Path Signature Anomaly} ($S_{\text{path}}$): Measures structural deviation from baseline path behavior, combining three sub-signals:
\vspace{-0.5em}
\begin{equation}
\small
S_{\text{path}} = w_l \frac{|\ell(P) - \mu_\ell|}{\sigma_\ell} + w_s \big(1 - J\big(\text{ngrams}(P),\, \text{ngrams}_{\text{base}}\big)\big) + w_b \cdot \mathbb{I}(b_{\text{new}})
\end{equation}
\vspace{-0.5em}

\noindent where $P$ denotes the critical path of the execution under evaluation, $\ell(P)$ is its path length (number of functions), $\mu_\ell$ and $\sigma_\ell$ are the mean and standard deviation of path lengths observed in the baseline executions, $J(\cdot)$ is the Jaccard similarity over bi-gram and tri-gram function name sequences, and $\mathbb{I}(b_{\text{new}})$ indicates whether the bottleneck function was unseen in baseline paths. The \emph{bottleneck function} is the function on the critical path with the highest individual wall-clock execution time; it is identified during critical path extraction (Section~\ref{sec:critical-path-extraction}). Default weights are uniform ($w_l = w_s = w_b = 1/3$), giving equal importance to the three sub-signals. The Jaccard term over function-name n-grams captures \emph{structural reordering} of function calls, which is a signal that neither path-length deviation nor the bottleneck indicator can detect.

\textbf{Resource Anomaly} ($S_{\text{res}}$): Quantifies deviation in resource consumption using z-scores against baseline distributions:
\vspace{-0.5em}
\begin{equation}
S_{\text{res}} = \max\!\Big(\frac{x_{\text{cpu}} - \mu_{\text{cpu}}}{\sigma_{\text{cpu}}},\; \frac{x_{\text{mem}} - \mu_{\text{mem}}}{\sigma_{\text{mem}}},\; \frac{x_{\text{io}} - \mu_{\text{io}}}{\sigma_{\text{io}}}\Big)
\end{equation}
\vspace{-0.5em}

\noindent Flagged when any resource metric exceeds its baseline 95th percentile (z-score $> 3.0$).

\textbf{Archetype Distribution Anomaly} ($S_{\text{arch}}$): Evaluates whether the observed archetype assignment has shifted from the expected distribution using the chi-square distance between observed and baseline archetype frequency distributions.

\textbf{Performance Bounds Violation} ($S_{\text{bounds}}$): A binary signal that checks for violations of established statistical limits:
\vspace{-0.5em}
\begin{equation}
S_{\text{bounds}} = \frac{1}{2}\big(\mathbb{I}(d > p_{99}) + \mathbb{I}(\ell > \ell_{\max})\big)
\end{equation}
\vspace{-0.5em}

\noindent where $d$ is the path duration and $p_{99}$ is the 99th percentile of baseline durations.

These scores are combined into a final anomaly score $S_{\text{anomaly}}$:

\vspace{-0.5em}
\begin{equation}
S_{\text{anomaly}} = w_1 S_{\text{path}} + w_2 S_{\text{res}} + w_3 S_{\text{arch}} + w_4 S_{\text{bounds}}
\end{equation}
\vspace{-0.5em}

\noindent where weights sum to 1. To ensure generalizability and avoid overfitting, we determined these weights using only the baseline (non-regressed) training data (70\% split), explicitly excluding all test data. Balanced weighting ($w_1 = w_2 = w_3 = w_4 = 0.25$), giving equal importance to all four signals, is used as the default configuration. The anomaly threshold $\tau = 0.65$ is similarly determined using only baseline training data. An execution is flagged as a regression if $S_{\text{anomaly}} > \tau$.

We evaluate our approach using the remaining 30\% of normal executions (to measure false positive rate) and the executions with injected regressions (to measure true positive rate). Performance is quantified using standard classification metrics: Precision, Recall, F1-score, and AUC-ROC. We compare our multi-signal approach against several baseline detection methods that capture the core detection strategies commonly employed in the regression detection literature: P95 duration thresholding~\cite{jiang2009automated, malik2013automatic}, statistical process control (3-sigma rules)~\cite{nguyen2012automated}, Isolation Forest anomaly detection~\cite{liu2008isolation, zhang2019robust}, and One-Class SVM~\cite{scholkopf2001estimating, zhang2019robust}. We note that existing regression detection tools such as PerfJIT~\cite{chen2020perfjit} and PEASS~\cite{reichelt2019peass} operate at version-comparison granularity and require code-change information as input, which differs from our setting where regressions are detected within execution-level behavioral signals; the selected baselines therefore represent the underlying detection strategies of these approaches applied to our feature space.

\subsection{Automation and Practical Workflow}

Our methodology is designed as an end-to-end automated pipeline: all stages (data collection, critical path extraction, static analysis, archetype discovery, and regression detection) are fully automated, requiring only initial compile flags and optional one-time threshold calibration.

To apply our methodology to a new C/C++ application, a developer must: (1) compile the application with \texttt{-finstrument-functions} to enable application-level instrumentation; (2) implement a workload generator subclass specifying input parameters and command templates (using the provided abstract base class); and (3) run the automated pipeline entry point, which handles trace collection, critical path extraction, archetype assignment, and regression detection. The replication package includes all scripts and a step-by-step guide for extending the framework to new applications.

%% file: sections/results.tex
\section{Results and Evaluation}
\label{sec:results}

\subsection{Experimental Setup}
\label{sec:setup}
The studied programs were compiled and executed on a machine with an Intel Core i7-11700K processor (3.60\,GHz), 16\,GB of RAM, and a 1\,TB NVMe SSD. For application-level tracing, we used uftrace v0.18, and for system-level tracing we employed LTTng~v2.13. The data collection process spanned approximately 10 machine-days and produced over 350\,GB of trace data.

\subsection{Preliminary Analysis: The Static-Dynamic Gap}
\label{sec:preliminary-study}

Before presenting our archetype-based results, we check whether static code complexity can predict runtime performance criticality. If static analysis were sufficient, our dynamic approach would be unnecessary.

For each function $f$, we compute a static complexity score combining structural metrics extracted via srcML:

\vspace{-0.5em}
\begin{equation} \label{eq:static_complexity}
\small
\begin{split}
S_{\text{static}}(f) = &\; \alpha_1\log\big(1 + \text{LOC}\big) + \alpha_2\log\big(1 + C_{\text{cyclo}}\big) + \alpha_3 N_{\text{nest}} \\
 &\; + \alpha_4\log\big(1 + \lvert\text{calls}\rvert\big) + \alpha_5\mathbb{I}(\text{has\_io})
\end{split}
\end{equation}
\vspace{-0.5em}

\noindent with weights from prior studies~\cite{neuhaus2010evaluating}: $\alpha_2 = 0.3$ for cyclomatic complexity~\cite{mccabe1976complexity}, $\alpha_1 = \alpha_3 = \alpha_4 = 0.2$, and $\alpha_5 = 0.1$. This composite formulation provides a stronger test than any single metric. We also compute a dynamic criticality score:

\vspace{-0.5em}
\begin{equation} \label{eq:dynamic_complexity}
S_{\text{dynamic}}(f) = \sum_{i=1}^{N} \mathbb{I}(f \in \text{CP}_i) \cdot \frac{T_f^{(i)}}{T_{\text{total}}^{(i)}}
\end{equation}
\vspace{-0.5em}

\noindent where $N = 1{,}500$ executions per application, $\text{CP}_i$ is the critical path for execution $i$, $T_f^{(i)}$ is the time in $f$, and $T_{\text{total}}^{(i)}$ is the total path duration. We use Spearman rank correlation ($\rho$), as all distributions were non-normal (Shapiro-Wilk, $p < 0.001$).

\textbf{Static metrics are poor predictors of runtime criticality.} Table~\ref{tab:rq3_correlations} shows that correlations are weak across all six applications, with static metrics explaining only 10.4\% of performance variance on average. Even OpenSSL, where complex algorithms are expected to be critical, reaches only $\rho = 0.543$.

\begin{table}[h]
\centering
\caption{Weak Correlation Between Static Complexity and Dynamic Criticality. \textbf{Bold}: highest $\rho$; \underline{underline}: lowest $\rho$.}
\label{tab:rq3_correlations}
\small
\resizebox{\columnwidth}{!}{
\begin{tabular}{lrrrr}
\toprule
\textbf{Application} & \textbf{Spearman $\rho$} & \textbf{$p$-value} & \textbf{Var. Exp. ($\rho^2$)} & \textbf{Functions ($n$)} \\
\midrule
SQLite     & \underline{$-0.041$} & 0.103 & 0.2\% & 1,547 \\
OpenSSL    & \textbf{0.543} & $<$0.001 & \textbf{29.5\%} & 108 \\
Zstandard  & 0.336 & $<$0.001 & 11.3\% & 565 \\
FFmpeg     & 0.175 & $<$0.001 & 3.1\% & 2,615 \\
cURL       & 0.407 & $<$0.001 & 16.6\% & 744 \\
jq         & 0.136 & $<$0.001 & 1.8\% & 618 \\
\midrule
\textbf{Average} & \textbf{0.259} & --- & \textbf{10.4\%} & --- \\
\bottomrule
\end{tabular}
}
\end{table}

\textbf{14.7\% of functions exhibit misaligned behavior.} We classified functions into four quadrants based on their static and dynamic ranks. The two misaligned quadrants are: \emph{hidden bottlenecks} (low static complexity, high dynamic criticality; 7.0\%) and \emph{misleading complexity} (high static complexity, low dynamic criticality; 7.6\%). The remaining functions fall into the \emph{aligned} quadrants (high-high and low-low). For example, single-line wrappers like \texttt{curl\_easy\_perform} and \texttt{jv\_dump} rank among the most critical functions despite minimal complexity, as they orchestrate expensive callee operations. We note that our static analysis is intra-procedural; invocation frequency and I/O latency are inherently dynamic and cannot be determined statically, which explains why inter-procedural analysis would only partially narrow this gap. Misalignment is most severe in I/O-intensive applications (SQLite: 19.9\%, FFmpeg: 14.4\%) and lowest in CPU-bound OpenSSL (4.6\%), suggesting that the gap between static and dynamic perspectives is largest when runtime factors like I/O latency and invocation frequency dominate.

\begin{boxK}
    \textbf{Preliminary Analysis Takeaway} \\
    Static complexity metrics explain only 10.4\% of runtime performance variance on average. This confirms that dynamic, execution-driven analysis is essential, motivating the archetype-based approach evaluated in RQ1 and RQ2.
\end{boxK}

\subsection{RQ1: Universal Performance Archetypes}
\label{sec:rq1}

\begin{table*}[!t]
\centering
\captionsetup{justification=centering}
\caption{Summary of the 13 Discovered Performance Archetypes. \\
\textbf{Bold} text indicates the three universal patterns found in all six applications (100\% universality).}
\label{tab:rq1_archetype_summary}
\small
\begin{tabular}{clrrrrll}
\toprule
\textbf{ID} & \textbf{Behavioral Pattern} & \textbf{Paths (\%)} & \textbf{Avg Dur.} & \textbf{Avg Depth} & \textbf{Univ.} & \textbf{Memory} & \textbf{Resource} \\
\midrule
A0$^\dagger$ & Mid-depth subsystem initialization & 8,992 (11.3\%) & 2.49ms & 13.5 & 5/6 & Volatile & CPU \\
\textbf{A1} & \textbf{Top-level lifecycle management} & \textbf{10,505 (13.2\%)} & \textbf{5.49ms} & \textbf{6.3} & \textbf{6/6} & \textbf{Growth} & \textbf{CPU} \\
A2 & Concentrated pipeline setup & 6,759 (8.5\%) & 38.15ms & 10.1 & 4/6 & Growth & CPU \\
A3 & Rapid library registration & 9,090 (11.4\%) & 0.17ms & 14.2 & 4/6 & Growth & CPU \\
A4 & Program compilation and parsing & 10,935 (13.7\%) & 19.42ms & 7.0 & 1/6 & Growth & CPU \\
A5 & Hash-driven function registration & 907 (1.1\%) & 2.15ms & 11.3 & 3/6 & Volatile & CPU+Mem \\
\textbf{A6} & \textbf{Sustained pipeline execution} & \textbf{8,164 (10.2\%)} & \textbf{25.04ms} & \textbf{10.1} & \textbf{6/6} & \textbf{Growth} & \textbf{CPU} \\
A7$^\dagger$ & Shallow bootstrap dispatch & 4,037 (5.1\%) & 3.21ms & 5.3 & 5/6 & Volatile & CPU \\
\textbf{A8} & \textbf{Deep multi-layer orchestration} & \textbf{13,297 (16.7\%)} & \textbf{4.52ms} & \textbf{16.9} & \textbf{6/6} & \textbf{Growth} & \textbf{CPU} \\
A9 & Memory-intensive deep initialization & 200 (0.3\%) & 43.33ms & 11.9 & 2/6 & Volatile & CPU+Mem \\
A10 & I/O-concurrent initialization & 780 (1.0\%) & 13.72ms & 14.4 & 4/6 & Growth & CPU+I/O \\
A11 & Recursive schema resolution & 2,347 (2.9\%) & 2.64ms & 19.3 & 1/6 & Volatile & CPU \\
A12 & Heavy codec lifecycle management & 3,776 (4.7\%) & 71.67ms & 11.7 & 1/6 & Growth & CPU \\
\bottomrule
\end{tabular}
\\
\vspace{1mm}
\small{$\dagger$~indicates near-universal archetypes ($\ge$5/6 applications). \\ Memory: Growth = monotonic increase; Volatile = non-monotonic. Resource: dominant resource consumed.}
\\
\vspace{-3mm}
\end{table*}

\begin{figure}[!b]
\centering
\includegraphics[width=0.95\columnwidth]{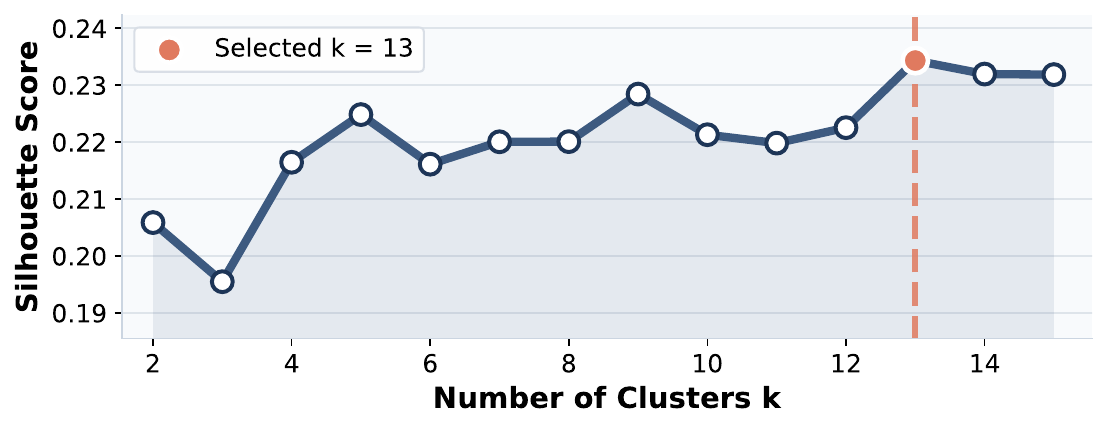}
\captionsetup{justification=centering,font=small}
\caption{Silhouette score vs.\ number of clusters $k \in [2,15]$.}
\label{fig:silhouette_vs_k}
\vspace{-4mm}
\end{figure}

\textit{RQ1: Can critical paths be automatically clustered into transferable performance archetypes?}
Following the clustering approach described in Section~\ref{sec:critical-path-extraction}, we applied $k$-means clustering to 79,789 critical paths extracted across six applications, each represented as a 33-dimensional feature vector.

\begin{figure}[t]
\centering
\includegraphics[width=0.95\columnwidth]{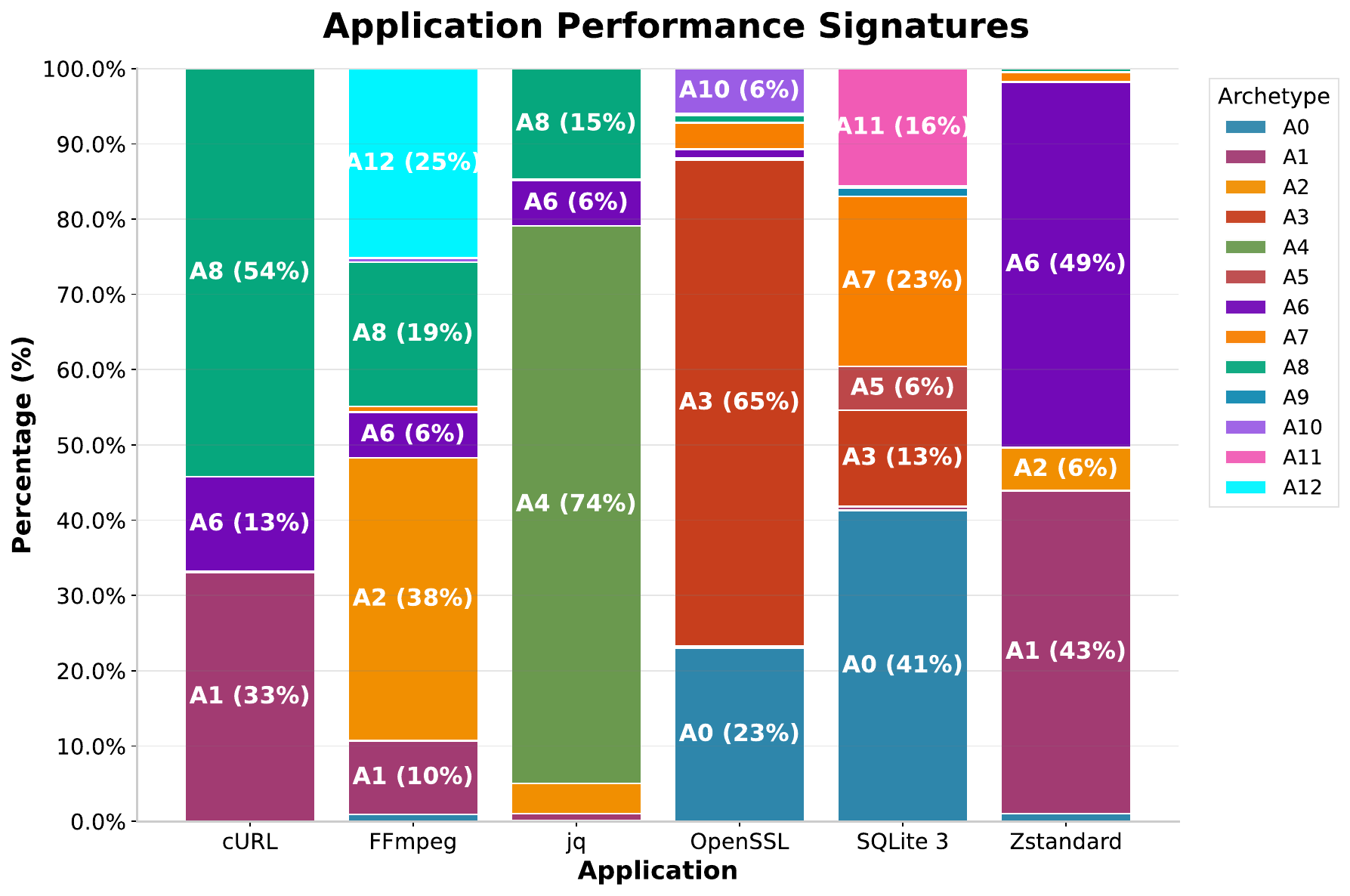}
\captionsetup{justification=centering,font=small}
\caption{Application Performance Signatures: Each bar shows how the performance archetypes contribute to a program's critical paths.}
\label{fig:rq1_signatures}
\vspace{-5mm}
\end{figure}

\textbf{Clustering quality and archetype count selection.}
We selected $k^*=13$ archetypes by evaluating silhouette scores for $k \in [2, 15]$ (see Figure~\ref{fig:silhouette_vs_k}). The selected configuration achieves a silhouette score of 0.235, Calinski-Harabasz index of 11,396.0, and Davies-Bouldin index of 1.340. The moderate silhouette score reflects the continuous nature of performance behavior; nevertheless, the resulting clusters produce semantically distinct, interpretable archetypes as shown in Table~\ref{tab:rq1_archetype_summary}.

\textbf{Clustering robustness across alternative paradigms.}
To verify that the archetypes are not artifacts of $k$-means, we ran a cross-paradigm check on a stratified subsample of 20,000 paths. Agglomerative-Ward at $k=13$ yields ARI\,=\,0.571 against the $k$-means partition (0\,=\,random, 1\,=\,identical), confirming the same groupings are recoverable hierarchically. HDBSCAN produces $\sim$22\% noise, consistent with known limitations of density-based methods in high-dimensional spaces~\cite{xu2005clustering}, with non-noise points agreeing moderately (ARI\,=\,0.250). These results confirm the 13 archetypes reflect genuine structure rather than algorithmic artifact.

\textbf{Despite diverse application domains, performance behaviors converge into 13 recurring archetypes.} Our clustering analysis identified 13 distinct archetypes that transcend application boundaries (Table~\ref{tab:rq1_archetype_summary}). To assign interpretive labels, we randomly sampled 50 critical paths from each cluster and examined their function call sequences, resource profiles, and behavioral characteristics, following standard clustering interpretation practices~\cite{spanos2008interpretation}. Three archetypes (A1, A6, A8) appear in all six applications, while two additional archetypes (A0, A7) appear in at least five. These five near-universal archetypes account for 44,995 paths (56.4\% of all observed paths), showing that over half of critical path behavior follows transferable patterns regardless of application domain.

\textbf{Call stack depth does not reliably predict execution duration.} We observed a striking disconnect between structural depth and runtime cost. For instance, Archetype A11 (average depth 19.3 functions) executes in only 2.64ms on average, while the much shallower A12 (depth 11.7) is the longest-running at 71.67ms on average. More dramatically, A3 (depth 14.2) completes in 0.17ms while A1 (depth 6.3) takes 5.49ms, which is a 32$\times$ difference inversely correlated with depth. The primary performance driver is the \emph{nature} of the operations (e.g., codec lifecycle management vs.\ library registration), not the structural complexity of the call stack.

\textbf{Each application exhibits a distinct ``performance signature.''}
Figure~\ref{fig:rq1_signatures} shows that each application's critical paths concentrate in characteristic archetypes. SQLite distributes across A0 (mid-depth subsystem initialization, 41.3\%), A7 (shallow bootstrap dispatch, 22.6\%), and A11 (recursive schema resolution, 15.6\%), reflecting its diverse query workload. OpenSSL is dominated by A3 (rapid library registration, 64.6\%), consistent with its many short cryptographic setup operations. Zstandard concentrates in A6 (sustained pipeline execution, 48.6\%) and A1 (top-level lifecycle management, 42.9\%), reflecting its streaming compression pipeline. FFmpeg distributes across A2 (concentrated pipeline setup, 40.0\%), A12 (heavy codec lifecycle management, 26.8\%), and A8 (deep multi-layer orchestration, 20.4\%), reflecting its complex multimedia processing pipeline. cURL concentrates in A8 (deep multi-layer orchestration, 54.2\%) and A1 (top-level lifecycle management, 33.0\%), consistent with its deep protocol stack and network I/O waiting patterns. jq is dominated by A4 (program compilation and parsing, 74.1\%), its sole occupant, reflecting that the majority of its execution is spent parsing and compiling JSON filter programs before execution.

Per-application diversity (measured as the fraction of the 13 archetypes observed in a given application) ranges from 0.31 (Zstandard, cURL) to 0.54 (SQLite, OpenSSL, FFmpeg). Zstandard and cURL's low diversity is consistent with their focused workloads (compression and network I/O, respectively), while SQLite's high diversity reflects its varied query processing patterns. jq exhibits moderate diversity (0.38), with its paths concentrated primarily in a single application-specific archetype (A4, 74.1\%).

\textbf{Application-specific archetypes are rare; most patterns generalize.} Only three archetypes are confined to a single application: A4 (program compilation and parsing, jq), A11 (recursive schema resolution, SQLite), and A12 (heavy codec lifecycle management, FFmpeg), collectively representing 21.4\% of paths. The remaining 10 archetypes (76.9\%) appear in at least two applications, showing high transferability of performance patterns. This suggests that optimization strategies developed for one application's dominant archetypes may benefit others sharing the same patterns.

\textbf{Near-universal archetypes enable transferable optimization guidance.} The five archetypes appearing in $\ge$5 applications map to concrete, transferable optimization strategies:
\begin{itemize}[topsep=2pt,itemsep=0pt,parsep=0pt,leftmargin=*]
\item \textbf{A8 (Deep multi-layer orchestration, 16.7\%):} The most frequent archetype, having deep call chains (16.9 depth) with moderate duration (4.52ms), representing multi-layer protocol and subsystem coordination. Reducing call overhead, inlining hot paths, and flattening call chains can improve performance.
\item \textbf{A1 (Top-level lifecycle management, 13.2\%):} Shallow paths (6.3 depth) with moderate duration (5.49ms) and monotonic memory growth. These represent top-level entry points managing application lifecycle (e.g., setup, run, teardown). Reducing per-invocation overhead and streamlining lifecycle transitions are primary strategies.
\item \textbf{A6 (Sustained pipeline execution, 10.2\%):} Shallow, moderate-duration (25.04ms) paths with monotonic memory growth, representing sustained data processing in main pipelines. Optimization should target algorithmic efficiency and memory allocation patterns within processing loops.
\item \textbf{A0 (Mid-depth subsystem initialization, 11.3\%):} Medium-depth paths (13.5 depth) with volatile memory patterns, representing subsystem initialization and resource provisioning. Lazy initialization and reducing redundant setup across subsystems are primary targets.
\item \textbf{A7 (Shallow bootstrap dispatch, 5.1\%):} Very shallow (5.3 depth), fast (3.21ms), volatile-memory paths appearing in 5/6 applications. These represent lightweight bootstrap and dispatch operations amenable to batching and caching.
\end{itemize}

\begin{table*}[h]
\centering
\caption{Comparison of Regression Detection Methods and Baseline Techniques}
\label{tab:rq2_baselines}
\small
\renewcommand{\arraystretch}{1.0}
\resizebox{0.925\textwidth}{!}{
\begin{tabular}{lcp{6.5cm}p{4.5cm}}
\toprule
\textbf{Method Name} & \textbf{CP-Aware?} & \textbf{Description} & \textbf{Input Features} \\
\midrule
\textbf{Our Multi-Signal Method} & \textbf{Yes} & Combines 4 signals (path anomaly, resource anomaly, archetype anomaly, bounds violation). & Path duration/length, Resource avgs, Archetype ID, Bottlenecks \\
\midrule
\textbf{CP-Aware Baselines} & & & \\
Threshold~\cite{jiang2009automated, malik2013automatic} & Yes & Rule-based detection using the 95th percentile (p95) of the baseline path duration distribution. & Path Duration only \\
Statistical Control~\cite{nguyen2012automated} & Yes & Statistical Process Control (SPC) using 3-sigma limits ($\mu \pm 3\sigma$) of baseline duration. & Path Duration only \\
Isolation Forest~\cite{liu2008isolation, zhang2019robust} & Yes & Unsupervised anomaly detection (contamination=0.1) trained on path and resource features. & Path Duration/Length, Resource avgs (CPU, Mem, I/O) \\
One-Class SVM~\cite{scholkopf2001estimating, zhang2019robust} & Yes & Unsupervised One-Class Support Vector Machine (nu=0.1) learning the normal boundary of path behavior. & Path Duration/Length, Resource avgs (CPU, Mem, I/O) \\
\midrule
\textbf{Resource-Only Baselines} & & & \\
Resource Threshold~\cite{jiang2009automated, malik2013automatic} & No & Flags if any resource metric (CPU, Mem, I/O) exceeds its p95 baseline threshold. & Resource avgs (CPU, Mem, I/O) \\
Resource Isolation Forest~\cite{liu2008isolation, zhang2019robust} & No & Isolation Forest trained exclusively on resource consumption metrics, ignoring path structure. & Resource avgs \& peaks (CPU, Mem, I/O) \\
Resource One-Class SVM~\cite{scholkopf2001estimating, zhang2019robust} & No & One-Class SVM learning normal boundaries of resource usage patterns only. & Resource avgs \& peaks (CPU, Mem, I/O) \\
\bottomrule
\end{tabular}
}
\\
\vspace{-2mm}
\end{table*}

\textbf{Archetype stability across iterations.} We measure iteration stability as the fraction of inputs for which all three execution iterations are assigned to the same archetype. Across all applications, we observe a 43.4\% stability. The remaining 56.6\% of inputs have at least one iteration assigned to a different archetype, reflecting genuine runtime non-determinism (thread scheduling, cache state, I/O timing) rather than methodology instability.

\vspace{-1mm}
\begin{boxK}
    \textbf{RQ1 Key Takeaways} \\
    Thirteen distinct performance archetypes capture all observed behaviors across six applications. These archetypes show high generalizability: 10 of 13 (76.9\%) appear in at least two applications, and five among them are near-universal (appearing in at least five of the six applications), collectively covering 56.4\% of execution paths. Each near-universal archetype maps to specific, transferable optimization strategies, enabling cross-application performance characterization and optimization guidance.
\end{boxK}

\subsection{RQ2: Triangulated Regression Detection}
\label{sec:rq2}
\textit{RQ2: Can performance regressions be reliably detected by triangulating multiple signals?}
Following the multi-signal detection approach described in Section~\ref{sec:regression-injection}, we built a performance baseline model from 70\% of normal executions and evaluated our triangulated framework on the remaining 30\% of normal executions plus the regression executions with injected anomalies. We compared our triangulated framework against seven baseline techniques representing the principal detection strategies used in the performance regression literature (Section~\ref{sec:regression-injection}). Table~\ref{tab:rq2_baselines} details the specific configurations, algorithms, and input features for each method evaluated.

\textbf{Our multi-signal method achieves high accuracy, with an F1-score of 0.867 and an AUC of 0.923.} Our framework successfully identified 2,089 of the 2,525 regressions (True Positives) while only misclassifying 206 normal executions as regressions (False Positives). This results in a high precision of 0.910, indicating that when our model flags a regression, it is very likely correct. It also achieves a strong recall of 0.827, demonstrating its ability to catch the majority of true performance regressions.

\textbf{Critical-path awareness is essential for detection, improving F1-score by 60.4\% over resource-only methods.} As shown in Table~\ref{tab:rq2_comparison}, our complete, critical-path-aware method (F1: 0.867) substantially outperforms all baselines. Its primary advantage comes from its holistic view; Compared to the best-performing \textit{resource-only} baseline (which is unaware of critical paths, F1: 0.540), our method improves the F1-score by 60.4\%. Resource-only baselines deliberately exclude path features to isolate the CP-awareness contribution, making the 60.4\% gap a direct, controlled measure. This shows that resource metrics in isolation are insufficient. Furthermore, our method also outperforms simplified critical-path detectors (e.g., Threshold, F1: 0.697), proving the value of synthesizing \textit{multiple} signals rather than relying on just one.

\begin{table}[b]
\centering
\caption{Performance Regression Detection Comparison}
\label{tab:rq2_comparison}
\small
\resizebox{0.95\columnwidth}{!}{
\begin{tabular}{lrrrr}
\toprule
\textbf{Method} & \textbf{F1} & \textbf{Prec.} & \textbf{Recall} & \textbf{AUC} \\
\midrule
\textbf{Our Multi-Signal (w/ CPs)} & \textbf{0.867} & \textbf{0.910} & \textbf{0.827} & \textbf{0.923} \\
\midrule
\textbf{Critical-Path-Aware Baselines} & & & & \\
Threshold & 0.697 & 0.911 & 0.565 & 0.755 \\
One Class SVM (CP Features) & 0.671 & 0.514 & 0.964 & 0.525 \\
Isolation Forest (CP Features) & 0.643 & 0.778 & 0.548 & 0.695 \\
Statistical Control & 0.480 & 0.939 & 0.323 & 0.651 \\
\midrule
\textbf{Resource-Only Baselines (No CPs)} & & & & \\
One Class SVM (Resource) & 0.540 & 0.426 & 0.739 & 0.369 \\
Threshold (Resource) & 0.407 & 0.707 & 0.286 & 0.583 \\
Isolation Forest (Resource) & 0.394 & 0.575 & 0.300 & 0.539 \\
\bottomrule
\end{tabular}
}
\end{table}

\textbf{The framework excels at detecting structural anomalies and resource spikes.} All three injection types (Table~\ref{tab:regression_types}) produce \textit{resource\_spike} and \textit{duration\_increase} events, plus \textit{path\_elongation} and occasionally \textit{new\_function} when the injected code enters the critical path. The injected anomalies manifested as different observable behaviors. Our framework achieves near-perfect accuracy for clear, discrete events like \textit{resource\_spike} (99.6\%) or the introduction of a \textit{new\_function} (97.9\%). It is also highly effective at identifying structural changes like \textit{path\_elongation} (85.5\%) and \textit{duration\_increase} (79.1\%). Detection of subtle behavioral changes like an \textit{archetype\_shift} (53.3\%) proved more challenging. This is likely because archetype shifts represent out-of-distribution changes in one of the model's most informative features: when a regression causes an execution to shift to a different archetype, the baseline statistical models, which were trained on the original archetype distribution, encounter feature values outside their learned boundaries, making it harder to distinguish genuine regressions from natural archetype variability across inputs.

\begin{figure}[]
\centering
\includegraphics[width=0.85\columnwidth]{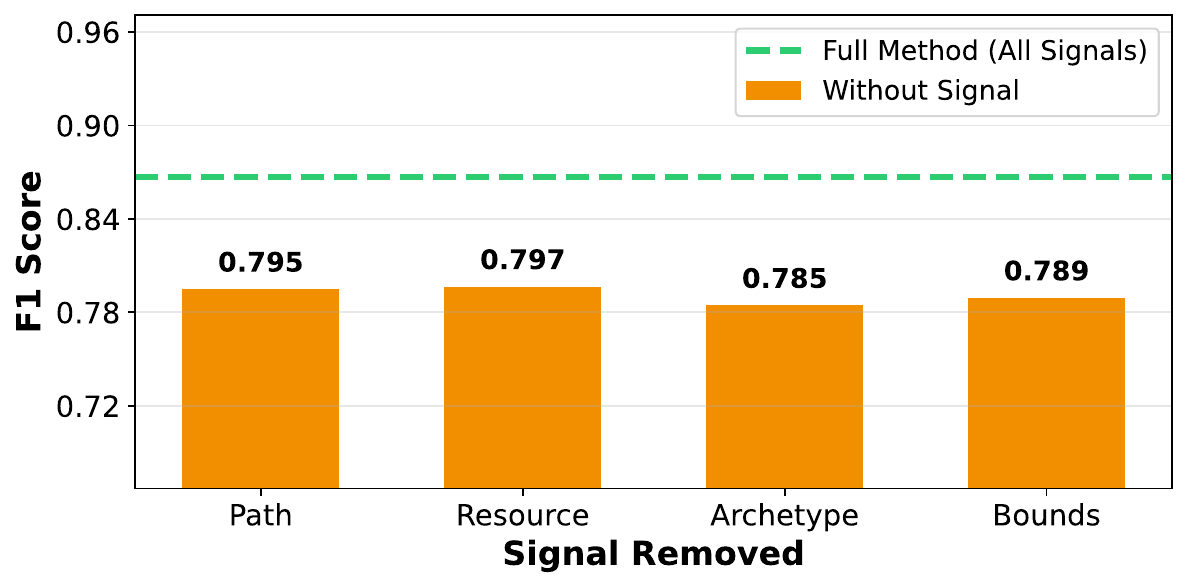}
\captionsetup{justification=centering}
\caption{Ablation analysis: signal contribution to detection performance regressions}
\label{fig:rq2_ablation}
\vspace{-3mm}
\end{figure}

\textbf{All four signals contribute meaningfully to detection accuracy.} To validate the necessity of our multi-signal approach, we performed an ablation study by systematically removing one signal at a time and measuring the impact on detection performance. Figure~\ref{fig:rq2_ablation} shows the results. The full method achieves F1: 0.867, while removing individual signals degrades performance: archetype distribution removal causes the largest drop (F1: 0.785, $-$9.5\%), followed by performance bounds (F1: 0.789, $-$8.9\%), path structure (F1: 0.795, $-$8.3\%), and resource consumption (F1: 0.797, $-$8.1\%). The consistent performance degradation when any signal is removed demonstrates that each perspective captures complementary information essential for robust regression detection.

\begin{boxK}
    \textbf{RQ2 Key Takeaways} \\
    Multi-signal triangulation achieves F1: 0.867, providing 60.4\% improvement over resource-only methods through critical-path awareness. The framework excels at detecting resource anomalies (99.6\% for resource spikes) and structural changes (97.9\% for new functions), and performs well on path elongation (85.5\%) and duration increases (79.1\%). Archetype shifts remain the most challenging to detect (53.3\%), as they represent out-of-distribution changes in the model's learned feature space. Reliable detection requires synthesizing structural, temporal, and resource perspectives rather than single-signal approaches.
\end{boxK}

%% file: sections/threats.tex
\section{Threats to Validity}
\label{sec:threats}

\subsection{External Validity}
Our study examines six open-source C/C++ applications, which may limit generalizability to other languages, programming paradigms, or proprietary industrial systems. However, our conceptual framework is language-agnostic: the critical path extraction, archetype discovery, and regression detection components operate on function-level timing data, which can be obtained from Java (via JFR or async-profiler), Python (via cProfile or py-spy), and Rust (via uftrace, which natively supports Rust). The main adaptation required for other languages is the tracing tool; our pipeline's feature computation operates on Trace Compass and TMLL's unified call-graph API, so a tracer swap suffices. A systematic per-language archetype analysis is reserved for future work. All experiments were conducted on a single Intel Core i7-11700K platform. While function-level execution profiles are generally stable across x86 generations, absolute timings and resource patterns may vary on different microarchitectures (e.g., AMD, ARM). Future work should validate on additional hardware. The randomly generated inputs, while diverse, may not fully represent real-world production workloads.

\subsection{Internal Validity}
Our regression detection evaluation relies on synthetic performance degradations injected into functions appearing on critical paths, which may not fully capture the complexity of real-world regressions that often result from interaction effects, subtle algorithmic changes, or configuration shifts~\cite{chen2017exploratory}. We inject one regression type per modified build; real-world scenarios may involve multiple simultaneous regressions. Although we mitigated confounding factors through multiple iterations (3 per input) and isolated environments, system-level noise may still introduce measurement variance. We used a single 70/30 train-test split (by input, not by iteration, to prevent data leakage) rather than cross-validation; partition sensitivity could affect reported metrics. The manual inspection process for archetype labeling, while following established clustering interpretation practices~\cite{spanos2008interpretation}, introduces subjective judgment.

\subsection{Construct Validity}
Our static complexity score (Equation~\ref{eq:static_complexity}) represents one possible formulation among many, and alternative weightings or features (e.g., pointer complexity, template instantiation depth) might yield different results. The choice of $k=10$ critical paths per execution and $k^*=13$ archetypes, while justified through silhouette analysis (Section~\ref{sec:rq1}), involves inherent discretization of continuous performance phenomena. The definition of ``misaligned functions'' using quartile-based thresholds (Q1/Q4) is one of several possible categorization schemes. Our critical path algorithm selects paths based on longest single invocation chains, which may miss functions that are cumulatively important through many short invocations; the dynamic criticality score $S_{\text{dynamic}}$ partially addresses this but the two metrics may yield different function rankings. Our 33-dimensional feature vector does not explicitly model cache behavior or branch prediction patterns; future work could validate archetypes against developer-identified bottlenecks.

%% file: sections/conclusion.tex
\section{Conclusions}
\label{sec:conclusion}

Motivated by the gap between static code structures and dynamic execution patterns, we discovered 13 critical-path-aware performance archetypes by clustering nearly 80,000 execution paths across six C/C++ applications. These archetypes show high generalizability: 10 of 13 appear in multiple applications, and five are near-universal, collectively covering 56.4\% of observed paths. The critical paths are further leveraged to build statistical performance models for regression detection, which by triangulating path structure, resource consumption, and archetype deviations outperforms resource-only baselines by 60.4\% (F1: 0.867).
Our results suggest that runtime performance behavior is more structured and transferable than commonly assumed: recurring archetypes emerge across diverse application domains, and multi-signal analysis substantially improves regression detection over single-perspective approaches. These findings point toward archetype-informed optimization strategies that generalize beyond individual applications.
Future work could broaden language and domain coverage, incorporate finer microarchitectural signals, and validate archetype transferability on additional hardware platforms.

%% file: sections/data_availability.tex
\section{Data Availability}

All data and code used in this study are available in the replication package\footnote{\url{https://github.com/mooselab/performance-archetypes}}, which contains all the materials to reproduce the work.

%% file: sections/acknowledgement.tex
\section*{Acknowledgement}

We acknowledge support from the Natural Sciences and Engineering Research Council of Canada (ALLRP 597968 - 24); projet rendu possible grâce au Fonds de recherche du Québec (\#361973).